%% file: branon.tex
\documentclass[onecolumn,prd,superscriptaddress,showpacs,floatfix, 
preprintnumbers,nofootinbib,showkeys]{revtex4-1}  

\usepackage{amsmath,amsfonts}
\usepackage{epsfig,epsf}
\usepackage{mathrsfs}
\usepackage{graphicx}
\usepackage{latexsym}
\usepackage{amsmath}
\usepackage{amssymb}
\usepackage{textcomp}
\usepackage{amsbsy}

\usepackage{graphicx,caption}
\DeclareGraphicsExtensions{.png,.eps,.pdf}	
\usepackage{axodraw}
\usepackage[dvips]{color}
\usepackage{subcaption}
\usepackage[normalem]{ulem}

\definecolor{darkgreen}{cmyk}{1,0,1,0.4}

\usepackage{slashed}
\usepackage[normalem]{ulem}
\usepackage{setspace}
\usepackage{multirow}

\renewcommand{\baselinestretch}{1.5}
\renewcommand{\arraystretch}{1.5}
\def\beq{\begin{equation}}
\def\eeq{\end{equation}}
\def\barr{\begin{array}}
\def\earr{\end{array}}
\def\dis{\displaystyle}

\begin{document}

\title{Relaxing the $W'$ constraint on compactified extra-dimension}

\author{Mathew Thomas Arun}
\email{thomas.mathewarun@gmail.com}
\affiliation{Center for High Energy Physics, Indian Institute of Science, Bangalore-560 012, India}

\date{\today}

\begin{abstract}
It has been known for some time that the inclusion of brane fluctuations,
namely branons, helps in the suppression of KK-mode couplings to brane localised matter fields. 
In this paper we study the constraint on brane tension and compactification scale for such models using the results from the direct search of $W'$ at 13 TeV LHC, with an integrated luminosity of $36.1 fb^{-1}$, in the case for which branon forms the entire cold dark matter. Unlike the rigid brane scenario, where the compactification scale gets constrained to $\frac{1}{R} \gtrsim 5.2$ TeV, here we show that compactification scales as small as $\sim 1$ TeV are allowed for brane tensions of similar strength.
\end{abstract}

\maketitle

\input{intro.tex}
\input{effective}
\input{interaction}

\input{conclusion}
\appendix
\input{Appendix}

\section*{Acknowledgement}
The author would like to thank Shiuli Chatterjee for the help and discussion on dark matter.
The author acknowledges the support from the SERB National Postdoctoral
fellowship [PDF/2017/001350].
\bibliographystyle{unsrt}

\bibliography{reference}

\end{document}

%% file: intro.tex
\section{Introduction}
One of the primary motivations for the search of new physics beyond Standard Model (SM) is to resolve the
well-known gauge hierarchy/naturalness problem in connection with the
fine tuning of the Higgs mass against large radiative corrections,
the reason being the enormous difference between the Electroweak and Planck scales.
Among several proposals to address this problem, models with extra
spatial dimensions draw special attention. While one such proposal \cite{ADD} solves the hierarchy between the scales by assuming the bulk fundamental scale of the order of Electroweak scale, with the
apparent hierarchy generated from the volume of the extra-dimensional space, the other \cite{RS} uses a dual interpretation of composite Higgs by assuming a bulk $AdS_5$ geometry, compactified on a circle of a radius of the order of the Planck scale. Neither of these remarkable solutions are without caveats. The astrophysical \cite{astro} and {\it Large Hadron Collider} (LHC) \cite{atlas,cms} measurements have placed tight constraints which push these models to the brink of their naturalness. 

Apart from these two families of models envisaged to solve the theoretical problems faced by Standard Model, flat extra dimensional models with TeV scale compactification \cite{UED} and their derivatives are known to contain a plethora of phenomenological observables. A minimal version of such models with Standard Model fields allowed to propagate in the five-dimensional bulk have been facing challenges recently from the LHC \cite{UEDLHC1,UEDLHC2} and dark matter relic observations \cite{wmap,planck}. There has been significant effort \cite{mtadivya} in solving these problems by modifying the geometry to include strongly coupled gravity in the  bulk.

In all the above models, it has been assumed that the segment of the extra-dimension is protected by very rigid world end branes. Though a dynamical understanding \cite{rubakov1,rubakov2,george,davies} regarding the origin of these positive tension branes could be from topological defects (like kinks) that arise in spontaneous symmetry breaking of a real scalar field in the bulk, in the set-up of the above models an effective treatment \cite{sundrum,dobado} for the brane is sufficient. Since these branes support localisation of matter fields on to them, any such localised field necessarily couple universally to all the Kaluza Klein (KK)-modes, with masses $M_n = \frac{n}{R}$, of the bulk fields.
This, along with the recent direct searches for a heavy charged gauge boson decaying into a final
    state lepton accompanied by a large missing
    energy at 13 TeV
    LHC, with an integrated luminosity of $36.1 fb^{-1}$ \cite{ATLASW}, rules out the compactification scale ($\frac{1}{R}$) upto 5.2 TeV. 

Nevertheless, due to the absence of rigid bodies in relativistic mechanics, any kind of a brane necessarily fluctuates. These fluctuations or branons are the Nambu-Goldstone bosons (or pseudo-NG bosons as shown in the next section) of the spontaneously broken translational symmetry in the extra-dimension due to the presence of the brane. These branon modes are also understood \cite{bando} to have suppression effect on the gauge boson KK mode coupling with brane localised fields. 
It has been shown \cite{Cembranos:2006wu,cembranos} that these zero-modes, due to their particular interaction structure, are stable and  could be responsible for the dark matter (DM) content of the universe. They constitute a wide spectrum ranging from hot to cold DM depending on their mass and the brane tension.

In this paper, we aim to relax the bound from $W'$ direct searches on models with brane localised fermions by including the brane fluctuations. Along with satisfying the LEP observables, we will study the interaction of the KK-1 partner of the W-boson with brane localised fermions in the limiting case where the branon field forms the entire cold dark matter and constrain the compactification scale for given brane tension. In our analysis we will show that though a large portion of the parameter space gets ruled out, small compactification scale ($\frac{1}{R}) (\sim 1$ TeV) for a comparable brane tension (f) is still allowed. The result becomes better on increasing the cutoff scale, but only at the cost of naturalness. Since we work in the limit where domain walls are very thin in comparison with the compactification radius, we will assume $f \gtrsim \frac{1}{R}$.

We will start the paper by briefly describing the effective low-energy description of the brane and its fluctuation in section \ref{effective}, and compute the  interactions of $W$-boson KK-modes with the brane localised matter in section \ref{interactions}. We will, using the 13 TeV LHC data, constrain the model parameters, namely the brane tension and the compactification radius and present the results in section \ref{conclusion}.

%% file: effective.tex
\section{Effective description of the brane and its Goldstone bosons}
\label{effective}
In the following we will briefly review the effective theory understanding of the massive brane fluctuations. Though the origin of the brane is not well understood, topological defects arising from the breaking of the $Z_2$ symmetry of a scalar $\lambda \phi^4$ theory can form stable domain wall solutions that localise matter fields. For brevity, we will use the low-energy effective description to understand the brane fluctuations and their influence on the W gauge boson interactions. A realistic picture of the origin of the brane is given in Appendix.

To that end, we will consider a 3-brane, with coordinates $x_\mu, \, \mu =0,1,2,3$, embedded in a five dimensional bulk that is compactified on $M_4 \times S^1$ ($x_M, \, M=0,1,2,3,5$), with compactification scale $1/R < M_s$, where $M_s$ is the cutoff scale of the effective model. 
It is convenient to parametrize the brane as $Y^M = (x^{\mu}, Y(x))$ with induced metric
$g_{\mu \nu}= \partial_\mu Y^M \partial_\nu Y^N G_{MN}$,
where $ G_{MN}$ is the metric in the bulk and its action could be written as
\[
S= -\tau \int_{M_4}\sqrt{g} \, d^4x \ ,
\]
where $\tau = \frac{1}{4 \pi^2}f^4$ is the brane tension. This particular form of brane parametrisation takes care of the gauge degree of freedom of the Lagrangian under $x \rightarrow x'$ in addition to the usual bulk coordinate invariance.

In the event of the brane having been created at a certain point $Y(x) = Y_0$, its presence breaks all the isometries involving the direction of the compact $S^1$.
The excitations along that direction corresponding to the zero modes could be parametrised by the Nambu-Goldstone boson field $\omega(x)$. Or in other words,
\[
Y(x) = Y(Y_0,\omega(x)) = Y_0 + \frac{\partial Y}{\partial \omega }\Big |_{Y=Y_0} \omega(x) + {\cal O}(\omega^2) \ .
\] 
To understand these Nambu-Goldstone bosons or branons, that arise on the breaking of this isometry, let us assume a bulk metric of the form,
\beq
G_{M N} = \begin{pmatrix}\tilde{g}_{\mu \nu}(x_\mu,x_5) & 0 \\
                                       0 & -1 \end{pmatrix} \ .
\eeq
Expanding $\tilde{g}_{\mu \nu}(x_\mu,x_5)$ about $x_5=Y_0$, we get
\[
\sqrt{g} = \sqrt{\tilde{g}}\Big( 1- \frac{1}{2}\tilde{g}^{\mu \nu}\partial_\mu \omega \partial_\nu \omega + \frac{1}{4}\tilde{g}^{\mu \nu}\partial_Y^2 \tilde{g}_{\mu \nu} \omega^2+ ... \Big) \ .
\]
Using this, the effective action for the brane could be written as,
\beq
\barr{rcl}
S &=& \dis \int_{M_4} d^4 x \sqrt{\tilde{g}}\Big[-\tau + \frac{\tau}{2} \Big(\tilde{g}^{\mu \nu}\partial_\mu \omega \partial_\nu \omega -M^2 \omega^2 \Big) \Big]\ ,
\label{actionbranon}
\earr
\eeq
where $M^2 = \frac{1}{2}\tilde{g}_{\mu\nu}\partial_Y^2 \tilde{g}^{\mu \nu}$ is the mass term due to the explicit breaking.
If we had chosen a metric that is independent of the deformation, the branons would have remained massless. This is similar to the explicit chiral symmetry breaking by the quark masses, which leads to massive pions. Since the branons form pseudo-scalars on a 3-brane \cite{Cembranos:2006wu}, all Lagrangian terms with odd number of branon legs vanish under the parity symmetry. Hence, they are always stable and the massive ones are ideal candidates for cold DM \cite{cembranos}. 

Though the branon physics is interesting in its own regard, in this paper we are interested in the influence of branons on the gauge interaction with brane localised fields. From the above action, we could derive the 3+1 dimensional, position space, 2-point correlator for the pseudo-Goldstone boson in the time-like region to be \cite{zhang},
\beq
\label{scalarpropM}
\langle \omega(x)\omega(y)\rangle = S_\omega(x-y) = {\it i} \frac{1}{\tau} \frac{M}{8 \pi |x-y|}H_1^{(2)}(M \, |x-y|) \ ,
\eeq
where $H_1^{(2)}$ is the Hankel function of the second kind.
Note that in the limit $M\rightarrow0$, the above propagator becomes
\[
\langle \omega(x)\omega(y)\rangle|_{M\rightarrow 0} =- \frac{1}{\tau} \frac{1}{4 \pi^2 |x-y|^2} \ ,
\]
which is the well known massless scalar propagator and matches with the zero mode kink fluctuation as derived in the Appendix. 
In the next section, we will analyse the interaction of bulk gauge
bosons with the brane localised fermions.

%% file: interaction.tex
\section{Model and $W'$ interaction}
\label{interactions}
In this model, we assume that the matter fields are localised on the brane and the gauge fields are free to propagate in the bulk.         
This will not only reduce the effect of the higher dimensional operators causing FCNC, but will also allow a rich phenomenology at LHC.
The excursion of gauge fields in the bulk of a compactified manifold necessarily brings along the discretised set of KK modes,  which couple to the brane and the localised matter fields, depending on their boundary conditions.
For phenomenological purpose, we choose Neumann boundary condition at both $x_5 = 0$ and $x_5=\pi R$, ensuring that the gauge boson has a zero mode which will be identified with the Standard Model gauge boson.
With these boundary conditions, the bulk gauge field could be expanded in Fourier series as,
\beq
A_\mu(x_\nu,x_5) = \frac{1}{\sqrt{2 \pi R}}A_{\mu}^{(0)} (x_\nu)+  \frac{1}{\sqrt{\pi R}} \sum_{n = 1} A_{\mu}^{(n)} (x_\nu) cos(M_n x_5) \ ,
\eeq
where the masses of the KK-modes are given by $M_n=\frac{n}{R}$. Assuming that the gauge boson corresponds to the $SU(2)$ weak sector, the lightest of the KK-modes is identified with the Standard Model W-boson gauge field.

The Lagrangian term that corresponds to the interaction of these bulk gauge bosons with the fermion fields is given as,
\beq
g_5 \int d^4x dx_5 \bar{\Psi}(x_\mu) \gamma^{\mu} \Psi (x_\mu) \, \Big[\xi(x_5-\omega(x_\mu))\Big]^2 \, A_{\mu}(x_\mu,x_5) \ ,
\label{interactionxi}
\eeq
where $\Psi(x_\mu)$ is the brane localised fermions with the extra-dimensional wave profile $\xi(x_5-\omega(x_\mu))$.
We have derived the exact form of $\xi(x_5-\omega(x_\mu))$ for a fermion localised on the kink in the Appendix. 
Although, for simplicity and to arrive at a closed form analytic solution, we will work in the limit where the wave profile $\xi$ can be replaced by a delta function $\delta(x_5-\omega(x_\mu))$, it can be shown that the coupling thus obtained matches at a sub percent level (for a kink mass $m \gtrsim 10\, \frac{1}{R}$) to that obtained using the exact form.

Using the plane wave expansion for the gauge bosons, the interaction term above could be re-written as,
\beq
\barr{rcl}
\dis g_5 \int d^4x  \bar{\Psi}(x_\mu) \gamma^{\mu} \Psi (x_\mu) \, A_{\mu}(x_\mu,\omega(x_\mu))
& = & \dis g\int d^4 x \sum_n \bar{\Psi}(x_\mu) \gamma^{\mu} \Psi (x_\mu) \, A^{n}_{\mu}(x_\mu) cos( \frac{n}{R}\omega(x_\mu)) \\
& = & \dis g\int d^4 x \sum_n  e^{-\frac{1}{2} (\frac{n}{R})^2 S_\omega(M_s)} \bar{\Psi}(x_\mu) \gamma^{\mu} \Psi (x_\mu) \, A^{n}_{\mu}(x_\mu)\\
& =& \dis \int d^4 x \sum_n g_n \, \bar{\Psi}(x_\mu) \gamma^{\mu} \Psi (x_\mu) \, A^{n}_{\mu}(x_\mu)\ ,
\label{couplingdef}
\earr
\eeq
where $g = \frac{1}{\sqrt{\pi R}} g_5$. For each `n' (KK-level), an expansion of the cosine in the dynamical field $\omega(x_\mu)$ leads to an infinite series of even number of branon legs at the vertex. We obtain the second line after contracting all those branon legs leading to loops at the same point in space given by $S_\omega(M_s)$.
 Note that we have used a hard cutoff for the momentum as the effective theory is valid only upto scales smaller than $M_s$. Hence, the coupling of the `n$^{th}$' KK-mode of gauge boson with the brane localised fermion becomes,
\beq
\label{KKgf}
g_n = g \, e^{-\frac{1}{2} (\frac{n}{R})^2 S_\omega(M_s)} \ ,
\eeq

where, $S_\omega(M_s)$ is given as
\beq
S_\omega(M_s) = \frac{1}{\tau} \frac{M}{8 \pi l_s }H_1^{(2)}({\it i}M \, l_s) = \frac{1}{\tau} \frac{M \, M_s}{8 \pi }H_1^{(2)}({\it i}\frac{M}{M_s}) \ .
\label{Somega}
\eeq
In the above equation, we have replaced the $|x-y|$ in eq \ref{scalarpropM} with the corresponding Euclidian length scale, $l_s = M_s^{-1}$, which denotes the smallest length that could be probed due to the effective nature of the theory.
In the limit where $\frac{M}{M_s}\ll 1$, the above result matches with the expression given in \cite{bando} to the first order,
 \beq
 \label{KKgfM0}
g_n = g e^{-\frac{1}{2} (\frac{n}{R})^2S_\omega(M_s)}\Big|_{M\rightarrow 0} =  g e^{-\frac{1}{2} (\frac{n}{R})^2 \frac{M_s^2}{f^4}} \ .
\eeq 
Note that, in the limit where the brane is rigid, $f \rightarrow \infty$, the effect of the exponential factor in eq \ref{KKgf} becomes negligible. In this situation, the KK-1 partner of the W gauge boson couples to the brane localised fermions with Standard Model W-boson coupling strengths and hence, the experiment \cite{ATLASW} rules out compactification scales ($1/R$) up to $5.2$ TeV.

\begin{figure}[!h]
           {
               \begin{center}
                  \vspace*{-2pt}

\epsfxsize=8cm\epsfbox{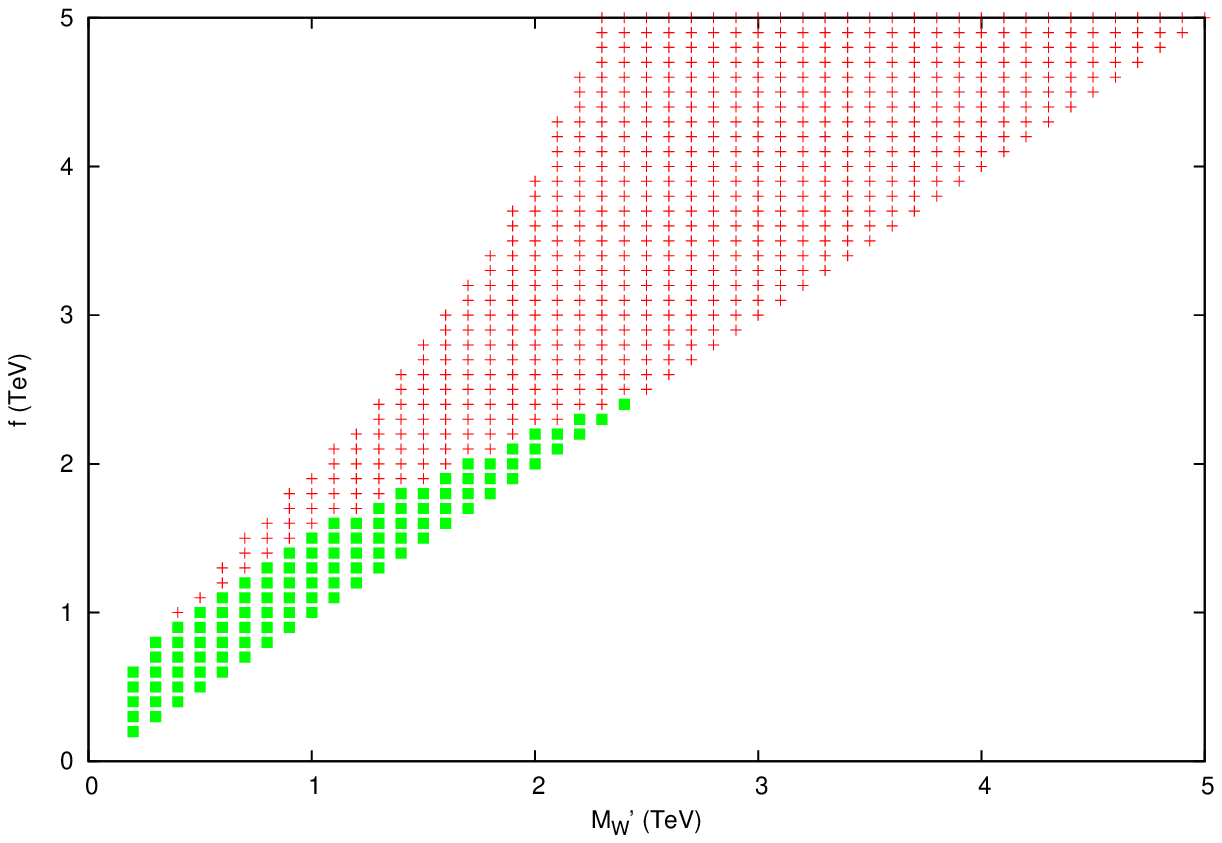}
\epsfxsize=8cm\epsfbox{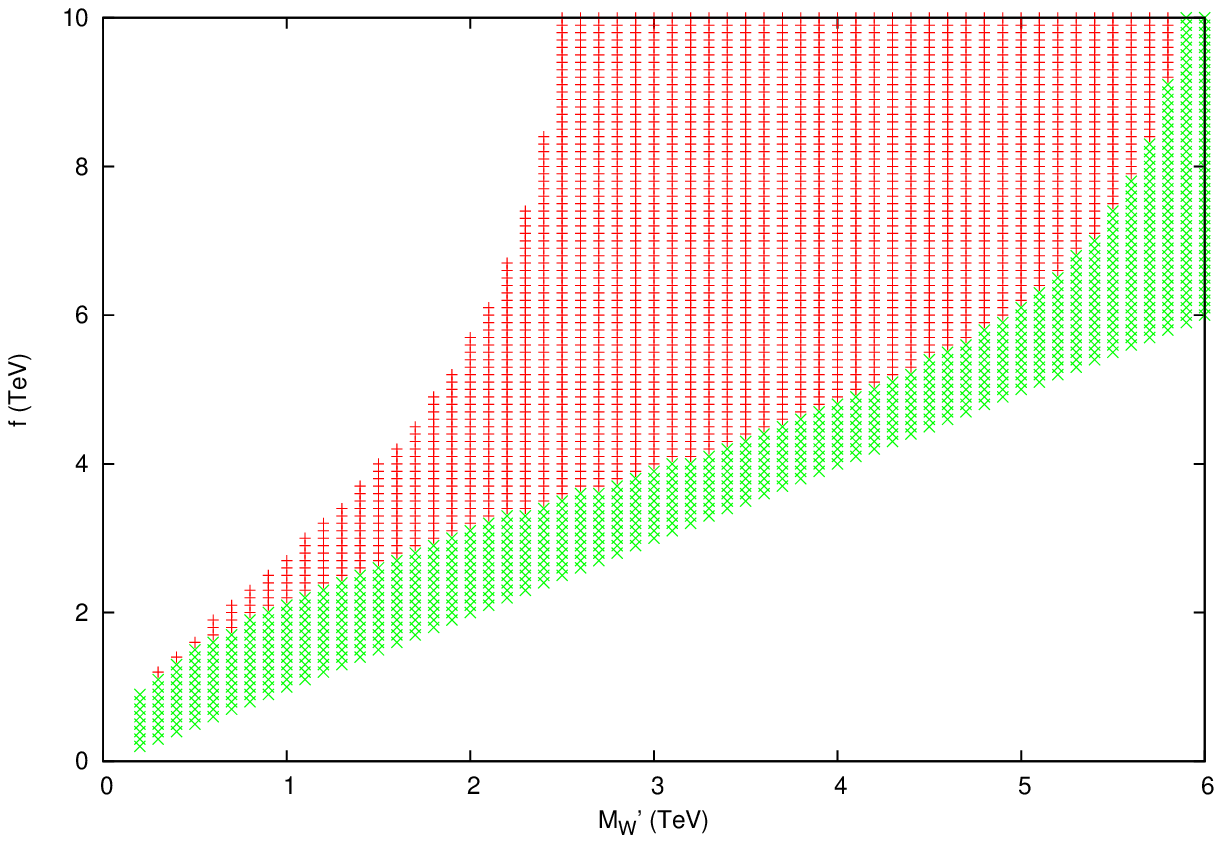}
\epsfxsize=8cm\epsfbox{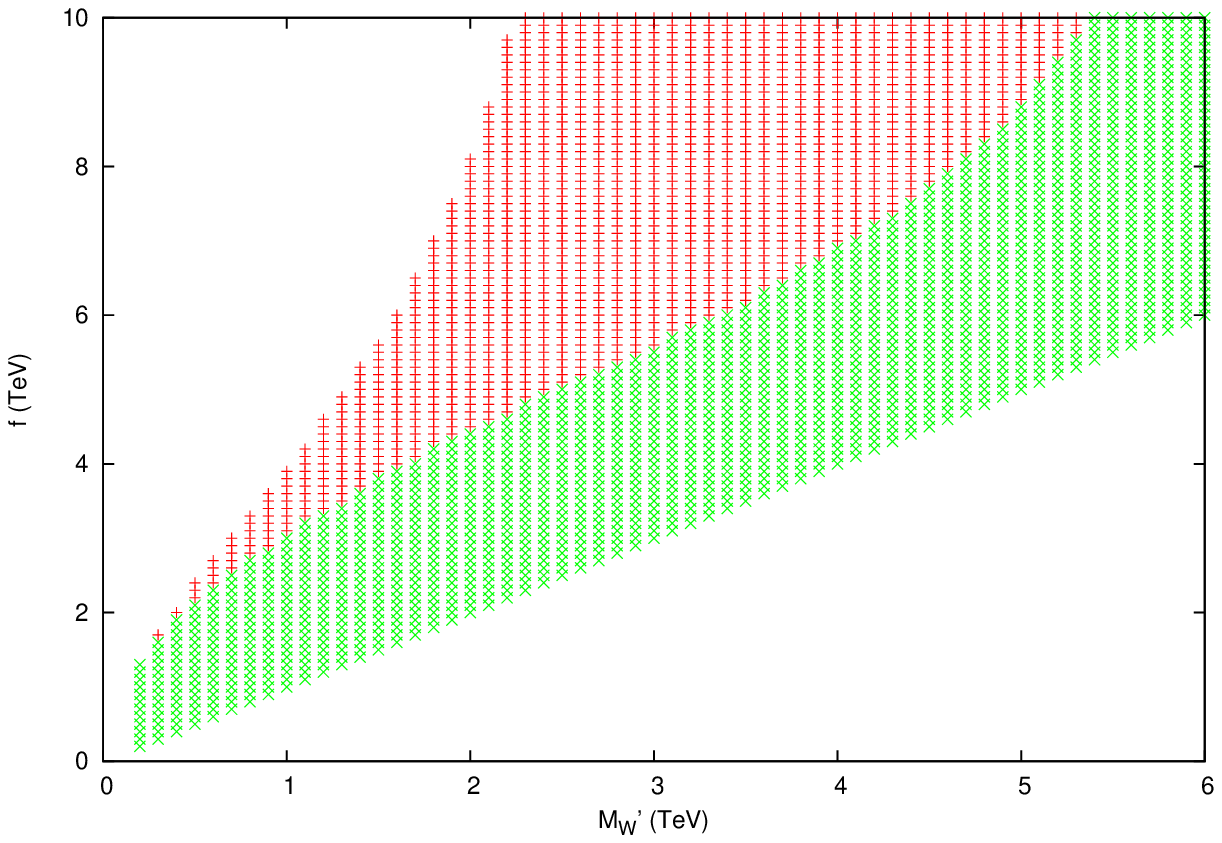}
\vspace*{-5pt}
\end{center}
}
\caption{The region, in the above plot showing brane tension vs compactification scale, disallowed by the W' and LEP searches is depicted in red (or $+$) and the allowed parameter space of the model is shown in green (or $\times$). We have kept the mass of branon a variable such that the dark matter relic density is always satisfied for a given brane tension. The figures corresponds to the cutoff scales $M_s =5$ TeV, $M_s=10$ TeV and $M_s = 20$ TeV respectively.}
             \label{fig:hankel}
\end{figure}

On the other hand, it is clear from eq \ref{KKgf} that for a finite tension brane the coupling of KK-1 state is smaller than that for the Standard Model W-boson, and hence the bounds could be relaxed. For that we need to know the mass of the branon for a given brane tension. If we demand that the annihilation cross-section of the branon be such that the observed relic density \cite{wmap,planck,cembranos} is reproduced, the mass of the branon and the brane tension get related. This branon mass corresponding to the brane tension, could be used in eq \ref{Somega} to compute the exponential factor in the coupling of KK-1 partner of the gauge boson with the fermions. With these couplings, we can compute the cross-section for the processes mediated by $W'$ that would contribute to LEP observables and at LHC. The most important contribution to the LEP observables comes from the correction to the four-Fermi operator (V). All the data points (both red (or $+$) and green (or $\times$)) shown in fig\ref{fig:hankel} are set to satisfy the constraint $V<0.0013$ at 95\% CL \cite{rizzo}. Then, we study the process $p \, p \rightarrow W' \rightarrow e \, \bar{\nu}_e$ at LHC using Calchep 3.7.4 \cite{calchep} and then compare this with the observed data \cite{ATLASW}. For a given KK-1 mode mass and cutoff $M_s$, the coupling that satisfies the LHC bound is identified at 95$\%$ CL, amongst the region that is constrained previously by the four-Fermi operator, and the corresponding parameter space is plotted as the green (or $\times$) region in fig \ref{fig:hankel}. Similarly, the region satisfied by the LEP observables but not by LHC is plotted as red (or $+$). 
Note that in all the plots, though a large section of the parameter space is unfavored, light KK-masses ($< 5.2$ TeV) are still allowed for small enough brane tensions.

%% file: conclusion.tex
\section{Conclusion}
\label{conclusion}
Extra-dimensional models, whether with flat large compactification radius or warped bulk, offer solutions to the theoretical challenges in Standard Model like fermion mass hierarchy, observed dark matter relic abundance, quadratic sensitivity of Higgs mass with new physics scale, smallness of the cosmological constant etc. Though they have interesting phenomenological observables, an alternative to Supersymmetric signatures at the {\it Large Hadron Collider} comes from the Universal Extra-dimensional scenario. In all of the above models and their derivatives a strict understanding of the origin of the brane is hard to come by. To simplify the involved computations in them it was sufficient to assume rigid branes. On the other hand, the rigidness of the branes comes with divergences which are not physical.

A dynamical explanation for field localisation on brane comes from the studies conducted by Rubakov and Shaposhnikov \cite{rubakov}, where they considered domain walls formed by kink fields \cite{coleman} with Yukawa coupling that lead to the trapping of matter in the potential well created by the kink. The lightest of the matter fields gets localised, whereas, the continuum which are heavier than the mass of the kink, escape into the bulk. The presence of these topological defects, on the other hand, break the translational symmetry in those directions. This manifests itself as the Goldstone bosons associated with the symmetry.

Rigid branes lead to a very simple, but unphysical picture. Including the brane fluctuations, as shown \cite{bando}, lead to the suppression of bulk gauge field couplings with brane localised matter, thus curing the brane-induced divergences present in the rigid brane models. It was also shown \cite{Cembranos:2006wu,cembranos} that these fluctuations form good candidate for a wide array of dark matter models, from hot to cold depending on their mass. 

In this paper we considered a generic five-dimensional model with gauge fields in the bulk and fermion zero modes localised to a finite tension brane. We constrained this model using the $p \, p \rightarrow W' \rightarrow e \, \bar{\nu}_e$ process while including the coupling suppression due to the branon field. Unlike the previous study \cite{bando}, where branons were assumed to be massless, in light of hot dark matter being ruled out, we considered massive branons that satisfy the cold dark matter relic and found that a large region of the parameter space, allowed by LEP, is unfavored by the 13 TeV LHC $W'$ direct searches at 95$\%$ CL. For small cutoff scales, the LHC data restricts the model to be favourable only in the parameter space with ${\cal O} \lesssim 1$ TeV brane tension and small compactification scale. The results become better when we assumed large cutoff scale, which theoretically is unfavourable from Higgs vacuum stability and U(1) perturbative arguments.

Though a majority of the region is ruled out, a ray of hope to find light KK-partners of bulk SM fields, at a high intensity collider,  is encouraging.

%% file: Appendix.tex
\section*{Appendix}
\label{Appendix}
Extra-dimensions, if present, should be hidden away since we have not observed them. This is possible, if, on integrating out the extra-dimensions, the extra degree of freedom manifests itself as infinite set of states called Kaluza-Klein modes and are very heavy. Hence, the central problem of model building is to device a mechanism that could provide with discrete states. The entrapment by the domain-wall solves precisely this problem in a manner that is fundamentally different from compactification. In this section we will briefly review the domain wall formation from a kink and its localisation properties. Afterwards, we will introduce collective coordinates, replacing the zero modes, to make manifest the Nambu-Goldstone bosons coming from the symmetry breaking of the translational degree of freedom. In this section, we use the notations given in \cite{george}.
\subsection{Kinks}
Let's consider a real scalar field $\Phi$ with a Lagrangian given by,
\beq
{\cal L} = \frac{1}{2} \, \partial^M \Phi \,\partial_M \Phi - V(\Phi) \ ,
  \qquad V(\Phi) = \frac{a}{4 \, m }\ \left[ \Phi^2- \frac{m^3}{a}
  \right]^2
  \label{potential}
\eeq
where $M=0,1,2,3,5$. While many different and inequivalent kink solutions are possible, note that it is the boundary conditions  that dictate the 
appropriate one. This is particularly true of the vevs at the fixed points. Assuming that the latter are so that the classical solution is nontrivial only along the $x_5$ direction, it is given by
\beq
\Phi_{cl}(x_{5})=\pm\sqrt{\frac{m^3}{a}} \,
            \tanh \left(\frac{m \, x_{5}}{\sqrt{2}}\right) \ .
\eeq
Henceforth, we refer to the positive (negative) sign as the 
kink (antikink) solutions. The energy of each is given by 
\beq
E_{\rm kink}=\int{dx_5 \frac{1}{2} [\partial_{x_{5}}\Phi_{cl}]^2+V(\Phi_{cl})}
=\int{d\Phi} \, [-2V(\Phi_{cl})]^{1/2}=\frac{2\sqrt{2} \, m^{4}}{3a} \ .
\label{energydensity}
 \eeq 
The modes about the kink solution can be obtained by 
effecting a perturbative expansion about it, namely
$\Phi(x^{M})=\Phi_{cl}(x_{5})+\tilde \phi(x^{M})$. Linearising 
the 
equation of motion for $\tilde \phi(x^{M})$, we have 
\[
\left[ \partial_M\partial^M - m^2 +\frac{3\, a}{m} \, \Phi^2_{cl} \right]
    \tilde \phi=0 \ .
\]
If we want to interpret this in terms of five-dimensional modes, 
we must re-express $\tilde \phi$ as 
\beq
\tilde \phi(x^{M}) = \Phi_{cl}(x^5 - \omega) + \sum_{i=0} \phi_{i}(x^\mu)
\, \eta_{i}(x_{5} - \omega)
\label{modeexpansion}
\eeq
where $\omega$ is a real number that denotes the position of the kink.
The equation of motion then simplifies to 
\[
\eta_i(x_5) \, \partial_{\mu}\partial^{\mu} \phi_i( x^{\mu})
    - \phi_i( x^{\mu}) \, 
      \left[ \partial_5^2 + m^2 - \, \frac{3\, a}{m} \, \Phi^2_{cl} 
        \right] \eta_i(x_5) =0
\]
Clearly, $\eta_i(x_5)$ has to be an  eigenfunction of the differential operator contained in the brackets. This is more conveniently expressed in terms of a rescaled dimensionless
 variable 
$z \equiv m \, (x_5-\omega) / \sqrt{2}$, to yield
\begin{equation}
\label{eq:kinkself}
\left[-\partial_z\partial_z-2+6 \, \tanh^2(z) \right]
 \eta_{i}(z)=\frac{2 \, M_{i}}{m^2} \, \eta_{i}
\end{equation}
Indeed, this is a particular example
of a general class of potentials $U_\ell(z) \equiv \ell \, (\ell + 1) \, 
\tanh^2 z - 2$. It should be noted that the one-dimensional Schr\"odinger equation is
invariant under a spatial translation of the field. The problem is well studied \cite{rajaraman} and the spectrum contains exactly $\ell$ discrete states with a continuum beyond. Though the
mode expansion given in eq \ref{modeexpansion} gave the energy spectrum
of the excitations of the kink field, the presence of zero-frequency
solution \cite{rajaraman} leads to divergences in the higher orders
unless treated on a different footing in comparison with the higher
modes. The method is to introduce `collective coordinates' to replace
the zero mode.
This is achieved by elevating the status of $\omega$ in
eq \ref{modeexpansion} to a field and re-writing the mode expansion of kink field as
\beq
\tilde \phi(x^{M}) = \Phi_{cl}(x^5 - \omega(x^\mu)) + \sum_{i=1} \phi_{i}(x^\mu)
\, \eta_{i}(x_{5} - \omega(x^\mu)) \ ,
\label{collectivemodeexpansion}
\eeq
where $\omega(x^\mu)$ is the 'collective coordinate' representing the
spatial position of the brane. Note that the zero mode is missing in
the above expansion and is taken care of by $\omega(x^\mu)$.
Now the Lagrangian given in eq\ref{potential} could be expanded as
\beq
{\cal L} = -E_{\rm kink} + \frac{1}{2} E_{\rm kink} \partial^{\mu}
\omega \partial_\mu \omega + {\rm higher \,  excitations} \ .
\eeq
Given the brane energy density, one could write it in terms of the brane tension as
$E_{\rm kink}=\frac{1}{4 \pi^2} f^4 $.
The configuration space propagator for the new zero mode becomes
\beq
S_\omega (x-y)= -\langle \omega(x)\omega(y)\rangle= \frac{1}{f^4}
\frac{1}{|x-y|^2} \ ,
\label{zeromodepropagator}
\eeq
where $f$ is the brane tension.

Let's now understand the  mechanism \cite{rubakov,george} to localize fermions on the kink.
Fermions in five-dimensions are Dirac spinors with four components, since the chiral operator does not exist. The Clifford algebra is spanned by $\Gamma^M$ such that $\{\Gamma^M,\Gamma^N\} = 2 \eta^{MN}$. We could choose the five-dimensional Dirac matrices to be $\Gamma^\mu = \gamma^\mu $ and $\Gamma^5 = - i \, \gamma^5$, where $\gamma^\mu,\gamma^5$ are the usual four-dimensional Dirac matrices. 
The action for a massless, five-dimensional fermion field $\Psi(x_\mu,x_5)$ coupled to a kink could be written as,
\[
S_\Phi = \int d^5x \Big( \bar{\Psi}i \Gamma^M \partial_M \Psi - \sqrt{\frac{ad^2}{2 m}} \Phi_{cl} \bar{\Psi}\Psi\Big) \ ,
\]
where $a$ and $m$ are as defined previously.

Expecting Lorentz symmetry breaking $M_5 \rightarrow M_4\times S_1$, five-dimensional fermion field could be expanded in terms of their four-dimensional spinor counterparts as,
\beq
\Psi(x_\mu,x_5) = \sum_i \psi_{L i }(x_\mu) f_{Li}(x_5) + \psi_{R i }(x_\mu) f_{Ri}(x_5) \ ,
\eeq
where $\psi_{L/R}$ are the four-dimensional chiral fermions given by $\gamma^5 \psi_{L/R} = \mp \psi_{L/R}$. Using the expansion above in the Dirac equation for the spinors, we get,
\beq
\psi_{L i }\Big( - \partial_5 f_{L i } + f_{Ri} M_i - \sqrt{\frac{a d^2}{2 m}} \Phi_{cl} f_{Li}\Big) +  \psi_{Ri} \Big( \partial_5 f_{R i } + f_{Li} M_i - \sqrt{\frac{a d^2}{2 m}} \Phi_{cl} f_{Ri}\Big) = 0 \ ,
\eeq
where we have used the four-dimensional Dirac equation $i\gamma^\mu\partial_\mu \psi_{L/R \, i } = M_i \psi_{R/L \, i} $. Since the above equation, on being separated, are not eigenvalue equations, we need to square them and turn them into eigenvalue equations of the form,
\beq
\barr{rcl}
\left[-\partial_z\partial_z + d(d+1)\, \tanh^2(z) -d \right] f_{L i}(z)& = &\frac{2 \, M_{i}}{m^2} \, f_{L i} \\
\dis \left[-\partial_z\partial_z + d(d-1)\, \tanh^2(z) +d \right] f_{R i}(z)&=&\frac{2 \, M_{i}}{m^2} \, f_{R i} \ .
\earr
\eeq
Again, we have used the substitution $z= m (x_5 -\omega)/\sqrt{2}$. 
As with the scalar fields, these are also an example of P{\"o}schl-Teller equation, and the solutions are well known. At the lowest mode, for $d=1$, we get one bound states with the five-dimensional wave profile
\beq
\barr{rcl}
M_0^2 =0 \qquad f_{L0}(x_5)&=& A_{L0} cosh^{-1}\Big(m (x_5 -\omega)/\sqrt{2}\Big)\\
\dis f_{R0}(x_5) &=& 0 \ ,
\earr
\eeq
where $A_{L0} $ is the normalisation constant.
In `collective coordinate' language, $\omega$ becomes a dynamical field and comparing to eq\ref{interactionxi} we see that $\xi(x_5-\omega(x_\mu)) =  A_{L0} cosh^{-1}\Big(m (x_5 -\omega(x_\mu))/\sqrt{2}\Big)$.